\documentclass[aps,prd,amsmath,amssymb,twocolumn]{revtex4}

\usepackage{graphicx}
\usepackage{dcolumn}
\usepackage{bm}
\usepackage{amsmath}
\usepackage{epstopdf}
\usepackage{amsfonts}
\usepackage{amssymb}
\usepackage{epsfig}
\usepackage{tabularx}
\usepackage{color}
\usepackage{soul}

\usepackage[colorlinks=true,citecolor=blue,urlcolor=magenta,breaklinks]{hyperref}

\newcommand{\be}{\begin{equation}}
\newcommand{\en}{\end{equation}}
\newcommand{\bea}{\begin{eqnarray}}
\newcommand{\ena}{\end{eqnarray}}

\begin{document}

\title{Orbits of light rays in (1+2)-dimensional Einstein-power-Maxwell gravity: Exact analytical solution 
to the null geodesic equations}

\author{
Grigoris Panotopoulos  {${}^{a}$
\footnote{
\href{mailto:grigorios.panotopoulos@ufrontera.cl}{grigorios.panotopoulos@ufrontera.cl} 
}
}
\'Angel Rinc\'on {${}^{b}$
\footnote{
\href{mailto:angel.rincon@ua.cl}{angel.rincon@ua.cl} 
}
%} 
%Il{\'i}dio Lopes  {${}^{a}$
%\footnote{
%\href{mailto:ilidio.lopes@tecnico.ulisboa.pt}{ilidio.lopes@tecnico.ulisboa.pt} 
%}
}
}

\address{
${}^a$ Departamento de Ciencias F{\'i}sicas, Universidad de la Frontera, Casilla 54-D, 4811186 Temuco, Chile.  
\\
${}^b$ Departamento de Física Aplicada, Universidad de Alicante, Campus de San Vicente del Raspeig, E-03690 Alicante, Spain.
}

\begin{abstract}
We study photon orbits within (1+2)-dimensional Einstein-power-Maxwell non-linear electrodynamics assuming a static and circularly symmetric background. An exact analytical solution to the null geodesic equations for light rays is obtained in terms of the Weierstra{\ss} function. We investigate in detail the impact of the photon energy, the electric charge of the black hole, and the integration constant (initial condition) on the shape of the orbits.
\end{abstract}

\maketitle

%%%%%%%%%%%%%%%%%%%%%%%
\section{Introduction}
%%%%%%%%%%%%%%%%%%%%%%

Light has always been of paramount importance in the history of Science. Indeed, over the years considerable
progress has been made observing the electromagnetic radiation reaching Earth from distant light sources. 
To mention just a few, the absorption spectra of chemical elements, the accidental discovery of the cosmic microwave background radiation by Penzias and Wilson \cite{PenWil}, and the bending of light during the total solar eclipse in 1919, are only some examples among many others. 
As far as gravitation is concerned, the study of orbits of light rays and/or massive test particles in certain fixed gravitational background is one of the principal ways to explore the physics of a given gravitational field.
For instance, in the case of Einstein's General Relativity (GR) \cite{Einstein:1916vd} 
and Schwarzschild geometry \cite{Schwarzschild:1916uq}, the explanation of the perihelion precession of the planet 
Mercury around the Sun \cite{deSitter:1916zza} as well as the bending of light \cite{deSitter:1916zza} during a (total) solar eclipse which took place in May of 1919 (see the illustrative review \cite{Crispino:2019yew}) comprise two of the classical tests of GR \cite{tests}.

\smallskip 

Moreover, understanding how light propagates through space in the presence of massive bodies is critical to our understanding of the Universe, e.g. to characterize the nature of dark energy and dark matter. Indeed, understanding gravitational lensing of distant galaxies requires precise knowledge of how light bends near the strong gravitational 
field generated by galaxies. This is a central piece of any cosmological model \cite{M1}, such as the concordance 
$\Lambda$CDM model \cite{M2}.

\smallskip

Due to non-linearity and the mathematical complexity of non-linear partial differential equations, in most of the cases of interest analytic methods cannot work. Therefore most of the gravitational effects can be understood either i) numerically or ii) employing approximate methods. Obtaining exact analytic expressions, however, is always desirable at least for two reasons. The first one is that analytic expressions may serve as test beds for numerical methods, and they are also a good starting point for developing approximate approaches \cite{evaAIP}. The second reason is that a systematic 
study of all effects and of the complete parameter space may be performed only in the case of exact analytic expressions.
Moreover, it is only in this case that the structure and characteristics of all possible solutions, including the 
derivation of observable effects, such as the perihelion advance or light deflection, may be explored \cite{evaAIP}.

\smallskip

The orbits of light rays in fixed gravitational backgrounds of certain forms are described by
solutions to differential equations of elliptic or hyperelliptic type. The theory of those functions was 
studied long time ago by Jacobi \cite{Jacobi}, Abel \cite{Abel}, Riemann \cite{Riemann1,Riemann2} and 
Weierstra{\ss} \cite{Weierstrass}. A review of those achievements as well as a compact description of the 
complete theory may be found in \cite{Baker}. In particular, the motion of test particles in the 
Schwarzschild space-time was completely analyzed by 
Hagihara using elliptic functions back in 1931 \cite{Hagihara}. In recent times, over the last 20 years or 
so, elliptic and hyperelliptic functions have been employed to obtain exact analytic solutions to the geodesic 
equations for well-known geometries, such as Schwarzschild-(anti) de Sitter 
\cite{Kraniotis:2003ig,Hackmann:2008zza,Hackmann:2008zz}, Kerr \cite{Kerr:1963ud} with a non-vanishing 
cosmological constant \cite{Kraniotis:2004cz,Hackmann:2010zz}, regular black holes 
\cite{Beato1,Beato2,Beato3,Leo1,Leo2,Garcia:2013zud}, and higher-dimensional 
space-times \cite{Hackmann:2008tu,RN5D}.

\smallskip

In recent years gravity in (1+2) dimensions 
\cite{Banados:1992wn,Banados:1992gq,Carlip:1995qv,Birmingham:2001dt,Koch:2016uso,Rincon:2018lyd}
has attracted a lot of attention for several different reasons. The absence
of propagating degrees of freedom, the mathematical simplicity, and the deep connection to Chern-Simons theory
\cite{Ana,Witten1,Witten2} are some of the reasons why it is interesting to study three-dimensional gravity. 
In addition to that, (1+2)-dimensional black holes are a good testing ground for the study of their four-dimensional
counterparts, since properties of (1+3)-dimensional black holes, such as horizons, Hawking radiation and black
hole thermodynamics, are also present in three-dimensional black holes.

\smallskip

A special attention has been devoted for several decades now to non-linear electrodynamics (NLE), which has already a 
long history, and it has been studied over the years in different contexts. To begin with, it is known 
that classical electrodynamics is based on a system of linear equations. However, when
quantum effects are taken into account, the effective equations become nonlinear. The first models go many 
decades back, when Euler and Heisenberg calculated QED corrections \cite{Euler}, while Born and Infeld managed to obtain
a finite self-energy of point-like charges \cite{BI} in the 30’s. Moreover, assuming
appropriate non-linear sources, which in the weak field limit are reduced to the usual Maxwell’s electrodynamics, one may generate a new class of Bardeen-like \cite{Bardeen,Borde} BH solutions \cite{Beato1,Beato2,Beato3,Leo1,Leo2,Garcia:2013zud,Bronnikov,Dymnikova,Hayward,Cisterna}.
Those solutions do have a horizon, while at the same time their curvature invariants, such as the Ricci scalar, are 
regular everywhere. This is to be contrasted to what happens in the case of the standard Reissner-Nordstr{\"o}m solution \cite{RN}. Finally, a toy model that generalizes Maxwell’s theory in a straightforward manner leads to the so called 
Einstein-power-Maxwell (EpM) theory \cite{Martinez,PR1,Gurtug,PR2,Unver,Xu,PR3,PR4,PR5,PR6,Pano,Panah:2022cay,Rincon:2021gwd,Gonzalez:2021vwp}, described by a Lagrangian density of the form $L(F) \sim F^k$, where $F$ is the Maxwell invariant, and $k$ is an arbitrary rational number. One of the advantages of EpM theory is that the nice conformal
properties of the four-dimensional Maxwell’s theory are preserved in any number of space-time dimensionality $D$, 
provided that the power $k$ is chosen to be $k = D/4$, as it is easy to verify that for this particular value the electromagnetic stress-energy tensor becomes traceless.

\smallskip

Within lower-dimensional gravity the orbits of light rights have been investigated in \cite{Gonzalez:2020kbv,Kazempour:2017gho,Panotopoulos:2021tkk}. In the present work we study for the first time the orbits of light rays in the presence 
of charged black holes in non-linear electrodynamics with Einstein's three-dimensional gravity. 
Our work is organized as follows: In the next section we briefly review the equations of motion for test particles 
within Einstein-power-Maxwell non-linear electrodynamics. In the third section we focus on null geodesics for light rays, and we obtain exact analytic solutions describing photon orbits in a fixed gravitational field for a given geometry. Finally, we close our work in the last section with some concluding remarks.

%%%%%%%%%%%%%%%%%%%%%%%%%%%%%%%%%%%%%%%%%%%%%%%%%%%%%%%%%%%%%%%%%%%%%%%%
\section{Background geometry and equations of motion for test particles}
%%%%%%%%%%%%%%%%%%%%%%%%%%%%%%%%%%%%%%%%%%%%%%%%%%%%%%%%%%%%%%%%%%%%%%%%

\subsection{Background geometry}

In what follows we shall briefly introduce the corresponding Einstein-power-Maxwell Lagrangian density, and we 
will re-obtain the background, i.e., the lapse function and the modified electric field. Both of them are now slightly modified by the inclusion of non-linear charged fields.

Let us start considering the well-known three dimensional action taking into account the Einstein-power-Maxwell theory 
with a negative cosmological constant $\Lambda$ 
\begin{equation}
I[g_{\mu \nu}, A_{\mu \nu}]=\int \mathrm{d^3}x \sqrt{-g}\left[ \frac{1}{2\pi }\left( R-2\Lambda
\right) -L\left( \mathcal{F}\right) \right]
\end{equation}
adopting geometrized units where $c=\hslash=8G=1$. The non-linear electromagnetic 
Lagrangian density is defined by $L\left( \mathcal{F}\right) =\left\vert \mathcal{F}\right\vert ^{3/4}$,
with $\mathcal{F}$ being the Maxwell invariant defined as usual to be,  
\begin{equation}
\mathcal{F}=F_{\mu \nu }F^{\mu \nu },
\end{equation}%
First we vary with respect to the gauge potential, $\mathbf{A}$, to obtain Maxwell equations
\begin{equation}
\mathbf{d}\left( ^{\star }\mathbf{F}L_{\mathcal{F}}\right) =0\rightarrow 
\mathbf{d}\left( ^{\star }\mathbf{F}\left\vert \mathcal{F}\right\vert
^{-1/4}\right) =0,
\end{equation}
where a $^{\star }$ denotes duality. 

Next we vary with respect to the metric tensor, $g_{\mu \nu }$, to obtain the field equations for gravitation
\begin{equation}
G_{\mu }^{\nu }+\Lambda \delta _{\mu }^{\nu }=\pi T_{\mu }^{\nu },
\end{equation}
where the energy-momentum tensor of the power Maxwell field is computed to be
\begin{equation}
T_{\ \nu }^{\mu }=\frac{1}{2}
\Bigl( 4\left( F_{\nu \lambda }F^{\ \mu \lambda
}\right) L_{\mathcal{F}}-\delta _{\ \nu }^{\mu }L
\Bigl) ,
\end{equation}%
or more precisely
\begin{equation}
T_{\ \nu }^{\mu } = \frac{1}{2} \left\vert \mathcal{F}\right\vert ^{3/4}  \left( 
3\frac{\left( F_{\nu \lambda }F^{\ \mu \lambda }\right) }{\mathcal{F}}%
-\delta _{\ \nu }^{\mu }\right) .
\end{equation}
Now, we shall assume an ansatz for our $(2+1)-$ dimensional line element, as follows
\begin{equation}
\mathrm{d}s^{2}=-f(r) \mathrm{d}t^{2}+ f(r)^{-1} \mathrm{d}r^{2}%
+r^{2}\mathrm{d}\theta ^{2}.
\end{equation}
The electric potential can be described as
\begin{equation*}
\mathbf{A}=A(r)dt,\text{ \ \ }
\end{equation*}%
where 
\begin{equation}
\mathbf{F=dA=}E(r) \mathrm{d}r \wedge \mathrm{d}t,
\end{equation}%
and its dual is
\begin{equation}
^{\star }\mathbf{F=}E(r)r \mathrm{d}\theta 
\end{equation}%
Thus, the corresponding invariant take the concrete form
\begin{equation}
\mathcal{F}=F_{\mu \nu }F^{\mu \nu }=-2E(r)^{2}.
\end{equation}%
Once we get the invariant $\mathcal{F}$ we now are able to obtain the electric field, namely:
\begin{equation}
\mathbf{d}\left( E(r) r \mathrm{d}\theta \left[ 2E(r)^{2}\right] ^{-1/4}\right) =0,
\end{equation}%
and the reduced form of the electric field is
\begin{equation}
E(r)=\frac{\text{constant}}{r^{2}}.
\end{equation}
Taking advantage of the metric, and choosing the integration constant
proportional to the electric charge $Q$, the potential is obtained as
\begin{equation}
A(r)=-\frac{Q}{r}
\end{equation}
Rewriting the energy-momentum tensor conveniently we have
\begin{equation}
T_{\ \nu }^{\mu }=\frac{1}{2}\left\vert \mathcal{F}\right\vert ^{3/4}\text{diag%
}\left( \frac{1}{2} ,\frac{1}{2} , -1\right) 
\end{equation}%
Notice that the explicit form of $\mathcal{F}$ is given by 
\begin{equation}
\mathcal{F}=-\frac{Q^{2}}{r^{4}}
\end{equation}%
Please, keep in mind that  $Q$ is a constant related to the charge of the black
hole. 
In order to obtain the lapse function, we take the $tt$ component of Einstein
equations 
\begin{equation}
\frac{1}{2r}f^{\prime }\left( r\right) +\Lambda =\frac{1}{4}\pi 
\left\vert \mathcal{F}\right\vert ^{3/4}.
\end{equation}%
and we obtain, after the appropriated integration, that
\begin{equation}
f\left( r\right) = - M + \frac{r^{2}}{a^{2}} + \frac{\pi  Q^{3/2}}{2 r},
\end{equation}%
in which $M$ is the black hole mass and $\Lambda =-1/a^{2}$. We observe that the above-mentioned solution can 
be written utilizing the well-known BTZ black hole lapse function according to
\begin{align}
\begin{split}
f(r)  =  f_{\text{BTZ}}(r) + \frac{\pi  Q^{3/2}}{2 r},
\end{split}
\end{align}
where $f_{\text{BTZ}}(r)$ is the lapse function of the usual BTZ space-time given by
\begin{align}
f_{\text{BTZ}}(r) \equiv - M + \frac{r^{2}}{a^{2}}
\end{align}

\subsection{Geodesic equations for test particles}

The geodesic equation for electrically neutral test particles in a fixed three-dimensional circularly symmetric gravitational background of the form
\begin{equation}
\mathrm{d}s^2 = g_{tt} \mathrm{d}t^2 - g_{rr} \mathrm{d}r^2 - r^2  \mathrm{d} \theta^2 
\end{equation}
is given by \cite{Garcia:2013zud}
\begin{equation}
\frac{d^2x^\mu}{ds^2} + \Gamma^\mu_{\rho \sigma} \frac{dx^\rho}{ds} \frac{dx^\sigma}{ds} = 0
\end{equation}
where $s$ is the proper time, while the Christoffel symbols $\Gamma^\mu_{\rho \sigma}$ are computed by \cite{landau}
\begin{equation}
\Gamma^\mu_{\rho \sigma} = \frac{1}{2} g^{\mu \lambda} \left( \frac{\partial g_{\lambda \rho}}{\partial x^\sigma} + \frac{\partial g_{\lambda \sigma}}{\partial x^\rho} - \frac{\partial g_{\rho \sigma}}{\partial x^\lambda} \right)
\end{equation}
and it is clearly symmetric in the indices $(\rho, \sigma)$. It should be mentioned that 
the geodesic equation consists of a set of second order differential equations. However, the problem may 
be simplified using the conserved quantities, which correspond to the first integrals of motion.
To do that, we recognize that for $\mu=1=t$ and $\mu=3=\phi$ the geodesic equations take the simple form
\begin{eqnarray}
0 & = & \frac{d}{ds} \left( g_{tt} \frac{dt}{ds} \right) \\
0 & = & \frac{d}{ds} \left( r^2 \frac{d\phi}{ds} \right) 
\end{eqnarray}
Thus, we can define the following quantities
\begin{equation}
E \equiv g_{tt} \frac{dt}{ds}, \; \; \; \; \; \; L \equiv r^2 \frac{d\phi}{ds}
\end{equation}
As always, the first conserved quantity, $E$, is the energy, whereas the second constant, $L$ is the angular momentum,

The third equation is the only non-trivial equation and corresponds to $\mu=2=r$ \cite{Garcia:2013zud}
\begin{equation}
\left( \frac{dr}{ds} \right)^2 = \frac{1}{g_{tt} g_{rr}} \: \left[ E^2 - g_{tt} \left( \epsilon + \frac{L^2}{r^2} \right) \right]
\end{equation}
which may be also obtained from \cite{Garcia:2013zud}
\begin{equation}
g_{\mu \nu} \frac{dx^\mu}{ds} \frac{dx^\nu}{ds} = \epsilon
\end{equation}
We have two well-defined cases: i) when $\epsilon = 1$ (for massive test particles), and ii) when $\epsilon = 0$ for light rays. For practical reasons, we will introduce the effective potential
\begin{align}
V^2_{\text{eff}} &= g_{tt} \left( \epsilon + \frac{L^2}{r^2} \right)
\end{align}
after which the equation of motion takes the well-known form \cite{Garcia:2013zud}
\begin{equation}
\left( \frac{dr}{ds} \right)^2 = \frac{1}{g_{tt} g_{rr}} \: \left[ E^2 - V^2_{\text{eff}} \right]
\end{equation}
Finally, the orbit is found obtaining $r$ as a function of $\phi$, which is found to be
\begin{equation}
\left( \frac{dr}{d\phi} \right)^2 = \left( \frac{dr/ds}{d\phi / ds} \right)^2  = \frac{r^4}{L^2 \: g_{tt} g_{rr}} \: \left[ E^2 - g_{tt} \left( \epsilon + \frac{L^2}{r^2} \right) \right]
\end{equation}
When the Schwarzschild ansatz is implemented (as in this case), the latter expression is reduced to be
\begin{equation} \label{Arr}
R(r) \equiv \left( \frac{dr}{d\phi} \right)^2 = \frac{r^4}{L^2} \: \left[ E^2 - g_{tt} \left( \epsilon + \frac{L^2}{r^2} \right) \right]
\end{equation}
Finally, it should be mentioned that the differential equation must be supplemented by the appropriate initial condition. Gor a given background geometry the shape of the orbits depends on $(E,L)$ as well as the initial angle $\phi_i$.

%%%%%%%%%%%%%%%%%%%%%%%%%%%%%%%%%%%%%%%%%%%%%%%%
\section{Exact analytic solution for light rays}
%%%%%%%%%%%%%%%%%%%%%%%%%%%%%%%%%%%%%%%%%%%%%%%%

In this section, first we shall obtain the general expression for the orbits in terms of the Weierstra{\ss} function  \cite{Weierstrass}. After that we shall display graphically the orbits in the $(x-y)$ plane for different values of the 
photon energy and different initial conditions.
Let us start by setting the concrete form of the metric potentials. They are written as follows 
\begin{align} \label{metric1}
g_{tt} &=f(r) =  - M + \frac{r^{2}}{a^{2}} + \frac{\pi  Q^{3/2}}{2 r}
\\
g_{rr} &= f(r)^{-1}
\end{align}
Although in principle any value of the power $k$ may be assumed, it is more natural to consider the case $k=3/4$,
as in this case the trace of the electromagnetic stress-energy tensor vanishes, which is the main motivation to study
EpM theory in dimensions other than four. Therefore, in the discussion to follow we set $k=3/4$.
%
%We recall that the above lapse function arises in several non-standard theories of gravity, such as Weyl gravity \cite{Mannheim:1988dj}, massive gravity \cite{deRham:2010kj,Panpanich:2018cxo}, and scale-dependent gravity  \cite{SD1,SD0,SD8,SD10} 
%at leading order in $\xi$.
%
%Focusing on photon orbits we set $\epsilon=0$, and
To make progress, let us make the change of variable $u=1/r$, and the corresponding equation for the trajectories $u(\phi)$ is then written as follows
\begin{equation}
\left( \frac{du}{d\phi} \right)^2 = P_3(u) = b_3 u^3 + b_2 u^2 + b_1 u + b_0  
\end{equation}
where the corresponding coefficients are computed to be
\begin{align}
b_3 &=-\frac{1}{2} \pi  Q^{3/2} \\
b_2 &= M \\
b_1 &= 0\\
b_0 &=\frac{E^2}{L^2}-\frac{1}{a^2}\\
\end{align}
Firstly, we perform a linear transformation of the form $u(\phi)=A y(\phi) + B$, where the coefficients
$\{A,B\}$ are given by
\begin{eqnarray}
A & = & \frac{4}{b_3} \\
B & = & - \frac{b_2}{3 b_3}
\end{eqnarray}
Such a trick is useful to obtain the solution in terms of the Weierstra{\ss} function, $\wp(\phi-\phi_{in};g_2,g_3)$. 
Thus, the initial equation $(du/d\phi)^2=P_3(u)$ is rewritten as follows  \cite{thesis}
\begin{equation}
\left( \frac{dy}{d\phi} \right)^2 = 4y^3-g_2 y-g_3
\end{equation}
Here the Weierstra{\ss} cubic invariants $g_2,g_3$ are found to be \cite{thesis}
\begin{eqnarray}
g_2 & = & \frac{1}{16} \left( \frac{4b_2^2}{3} - 4 b_1 b_3  \right) \\
g_3 & = & \frac{1}{16} \left( \frac{b_1 b_2 b_3}{3} - b_0 b_3^2 - \frac{2 b_2^3}{27}  \right)
\end{eqnarray}
Finally, the full solution in term of the Weierstra{\ss} function, $\wp(\phi-\phi_{in};g_2,g_3)$, requires the 
determination of the integration constant, $\phi_{in}$. Thus, imposing the initial condition $u(\phi_0)=u_0$ we can obtain 
\begin{equation} \label{main1}
%\boxed{
r(\phi) =\frac{1}{A y(\phi) + B}=\frac{1}{A \: \wp(\phi-\phi_{in};g_2,g_3) + B}
%}
\end{equation}
provided that the discriminant 
\begin{equation}
\Delta \equiv g_2^3 - 27 g_3^2
\end{equation}
does not vanish, since when $\Delta=0$ the case is singular \cite{Kraniotis:2003ig}. 

The interested reader may consult for instance \cite{lectures} for more details regarding the Weierstra{\ss} function and its properties. Moreover, for people familiar with the Jacobi elliptic functions, we remark in passing that the Weierstra{\ss} function may be expressed in terms of them, see e.g. \cite{elliptic}.

\smallskip

The formalism used up to now is valid for a family of background geometries in (1+2) dimensions, 
which are static, circularly symmetric solutions of Einstein's gravity. To get insights regarding the potential 
effect of the non-linear electrodynamics on the the geodesic structure of the solution,
we shall show below a set of figures both the effective potential for photons as well as the corresponding orbits, 
varying the energy, $E$, the initial angle, $\phi_{in}$, and the electric charge, $Q$. 

\smallskip

\begin{figure*}[ht!]
\centering
\includegraphics[width=0.32\textwidth]{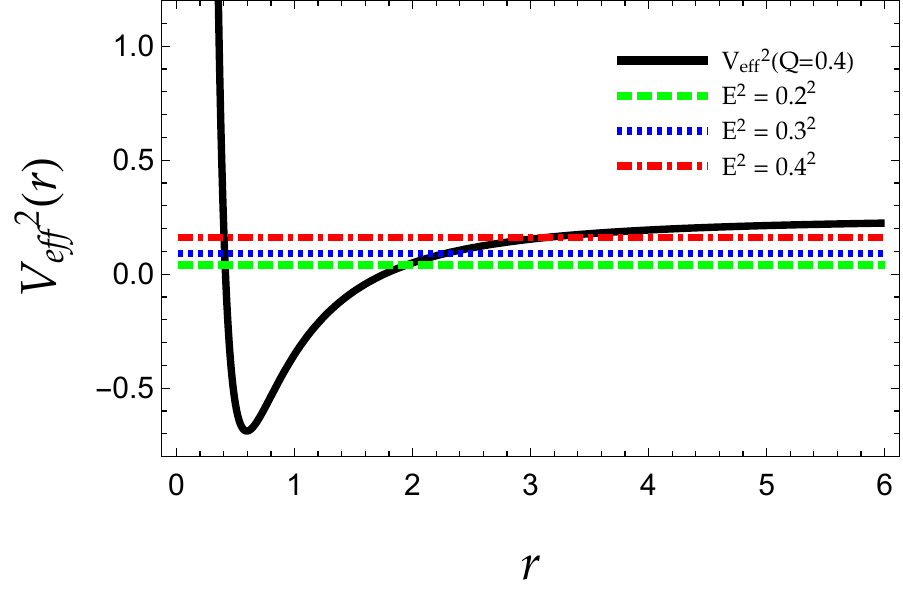}   
\includegraphics[width=0.32\textwidth]{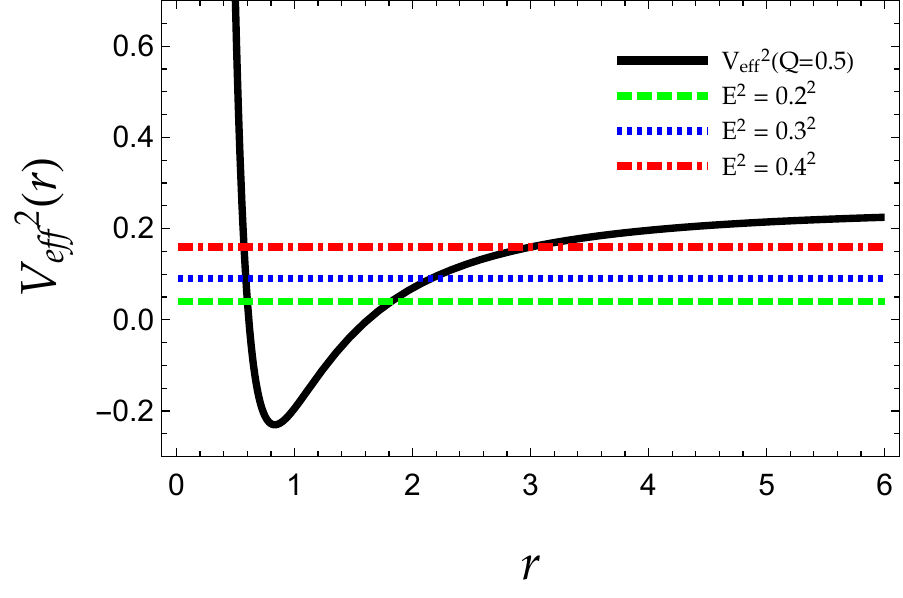}   \ \ 
\includegraphics[width=0.32\textwidth]{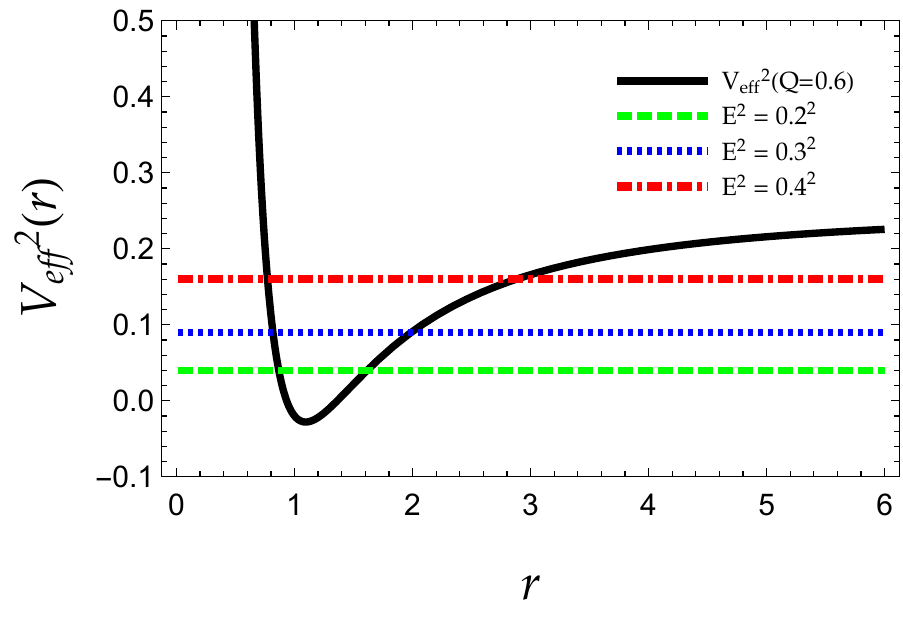}  % \ \
%\ \
%\\
%\includegraphics[width=0.32\textwidth]{Potential.pdf}   \ \
%\includegraphics[width=0.32\textwidth]{Potential.pdf}  
%}
\caption{
Effective potential for photons at different energy regimes in this geometry setting 
$M=1, a=2$ and $L=1$. We take three different values for the electric charge to evaluate the effective potential and three constant values for the energy.
}
\label{fig:potentialscase1}
\end{figure*}

\begin{figure*}[ht!]
\centering
\includegraphics[width=0.32\textwidth]{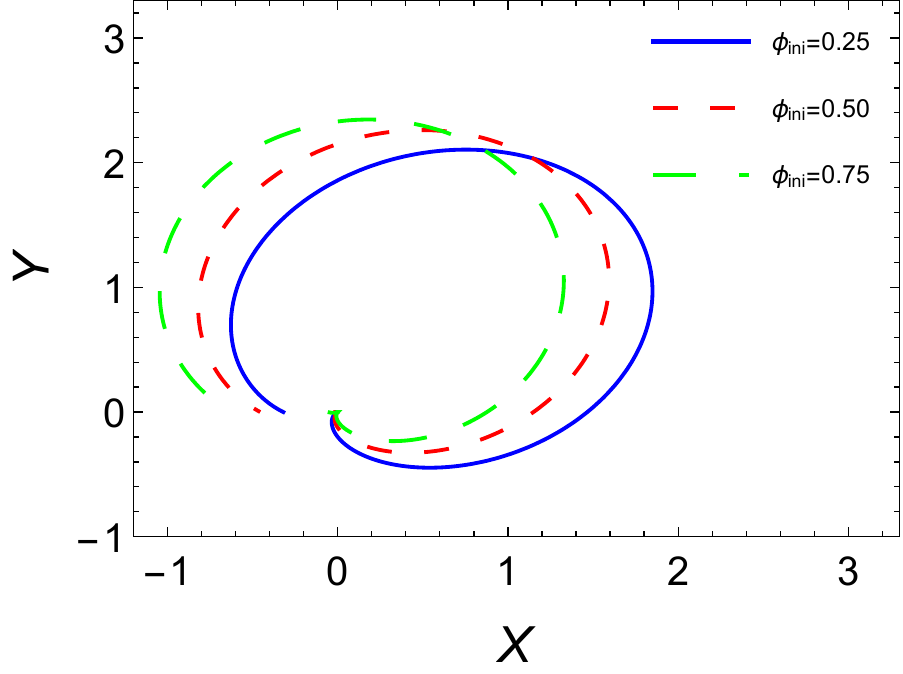}   \ \
\includegraphics[width=0.32\textwidth]{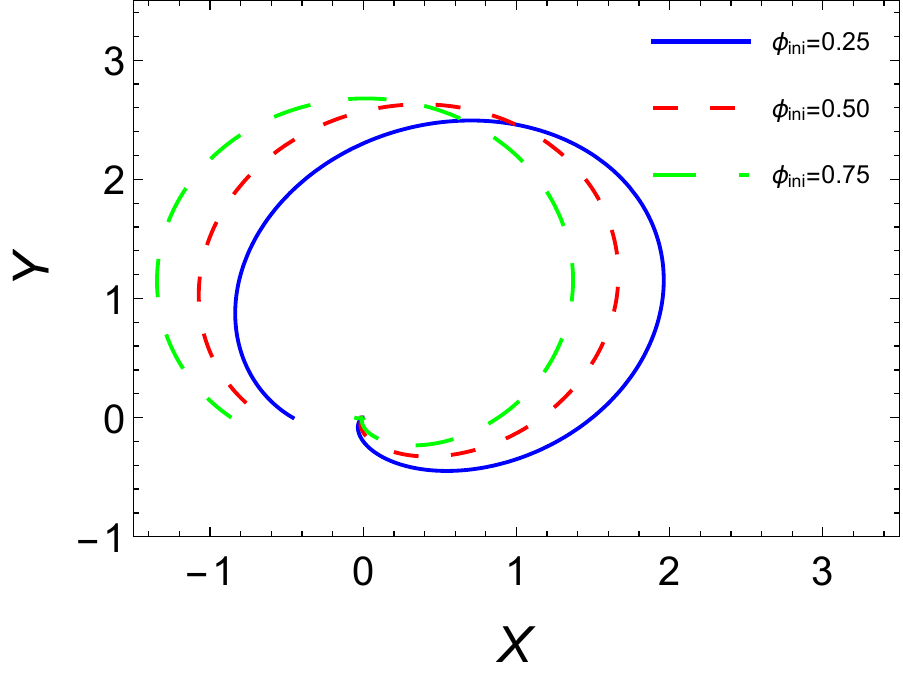}   \ \ 
\includegraphics[width=0.32\textwidth]{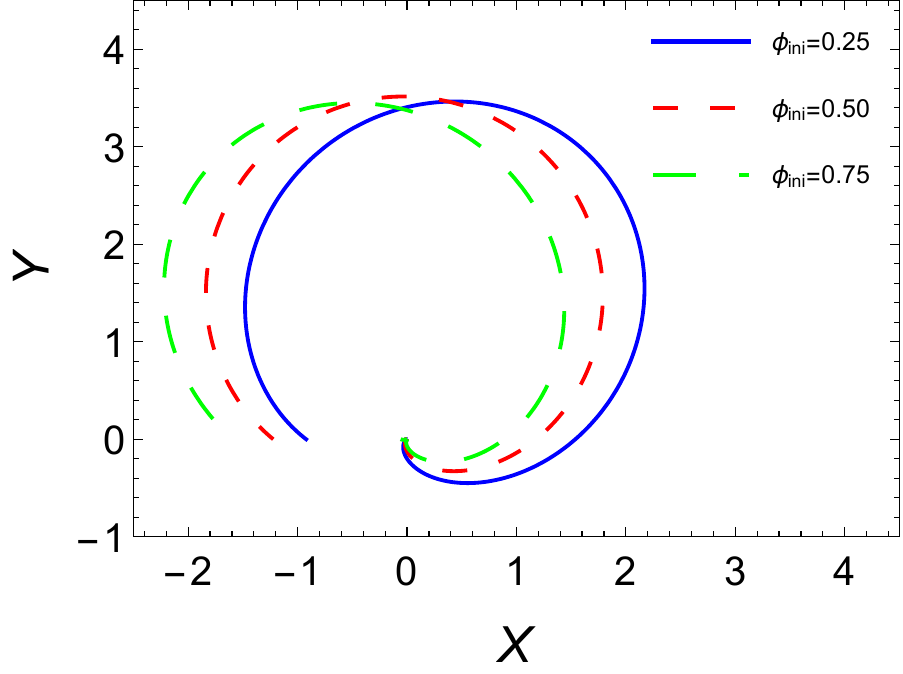}  % \ \
%\\
\includegraphics[width=0.32\textwidth]{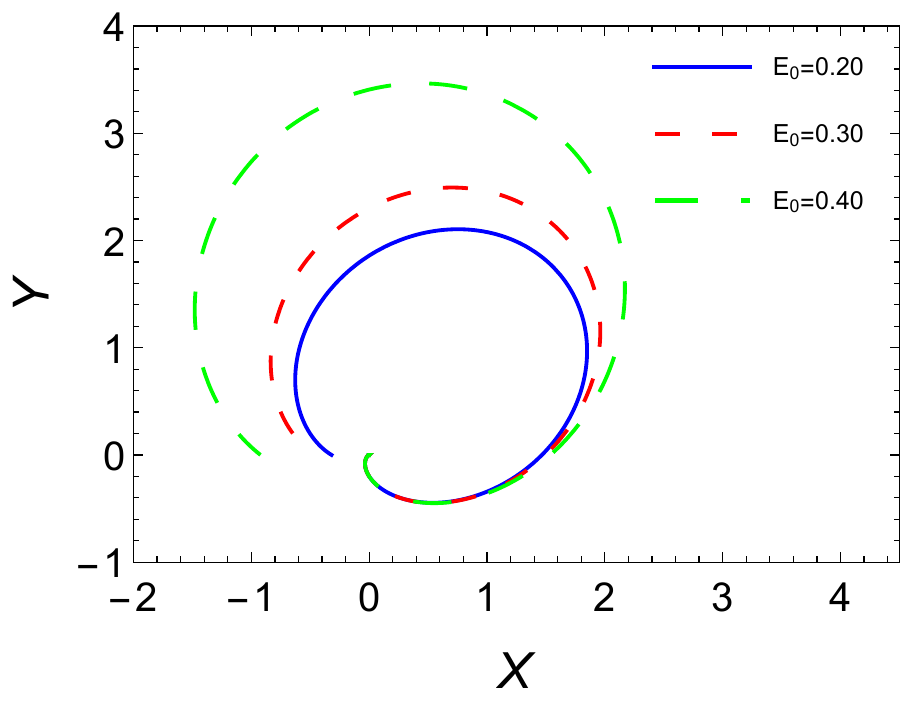}   \ \
\includegraphics[width=0.32\textwidth]{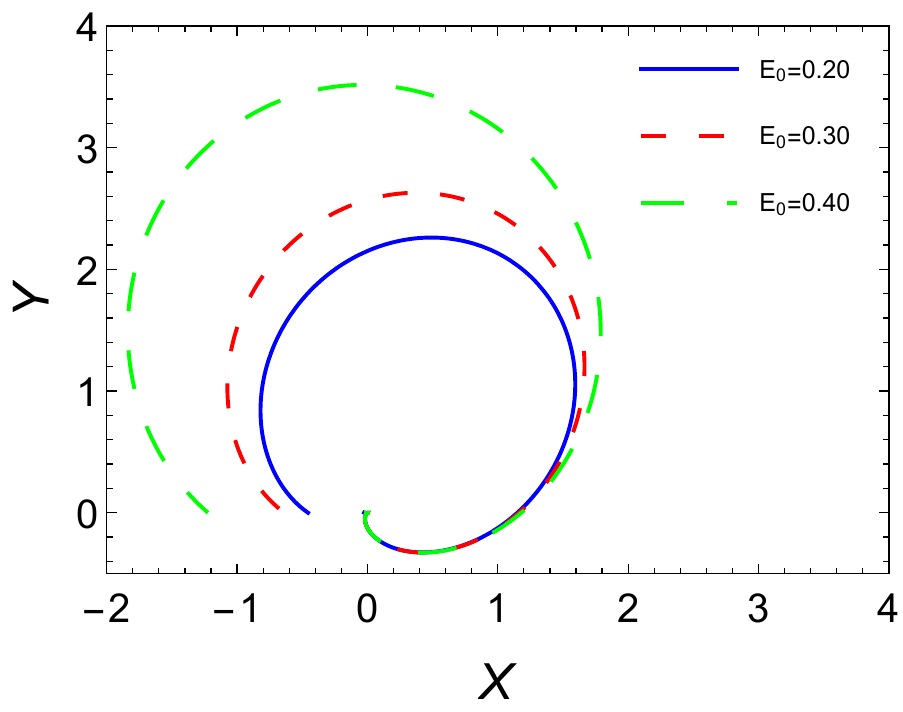}   \ \ 
\includegraphics[width=0.32\textwidth]{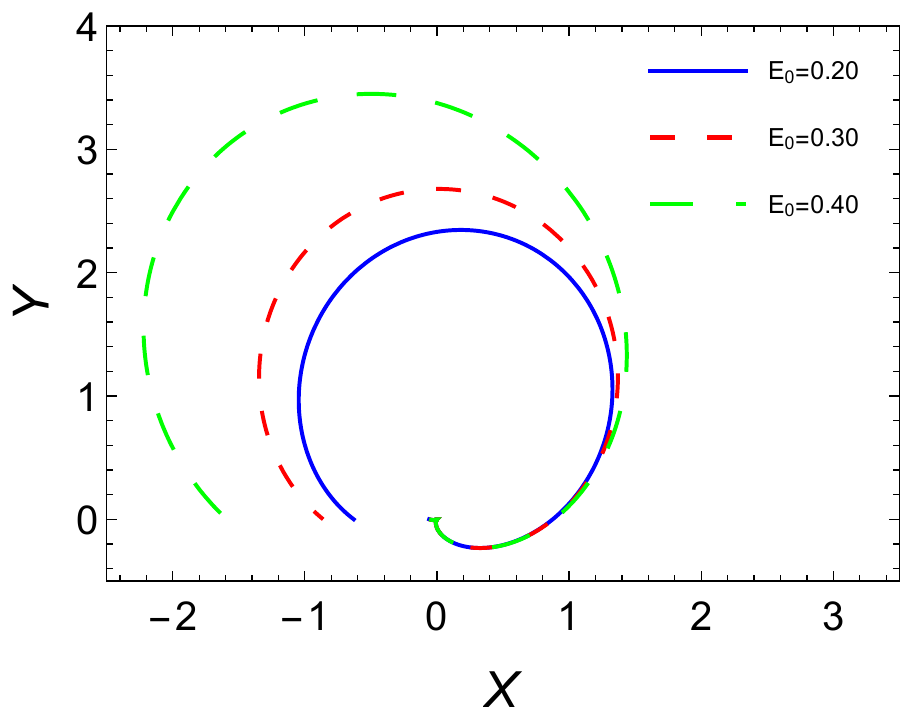}  % \ \
%\includegraphics[width=0.48\textwidth]{trayeSDE250v2.pdf}  
%}
\caption{ 
Photon orbits in the $(x-y)$ plane ($0 \leq \phi \leq 2 \pi$) in the Einstein-power-Maxwell geometry in (1+2) dimensions, setting $M=1, a=2, Q=0.4$ and $L=1$.
The impact of the photon energy and initial conditions is investigated
i) varying the initial angle for a given photon energy, and
ii) varying the energy for a given initial angle, as follows:
{\it{First row}}:
i) {\textbf{Left Panel:}} $E=0.20$ for different values of the initial angle $\phi_{ini}$.
ii) {\textbf{Middle Panel:}} $E=0.30$ for different values of the initial angle $\phi_{ini}$.
iii) {\textbf{Right Panel:}} $E=0.40$ for different values of the initial angle $\phi_{ini}$.
{\it{Second row}}:
i) {\textbf{Left Panel:}}  $\phi_{ini} = 0.25$  for different values of the energy $E$.
ii) {\textbf{Middle Panel:}} $\phi_{ini} = 0.50$   for different values of the energy.
iii) {\textbf{Right Panel:}} $\phi_{ini} = 0.75$  for different values of the energy.
}
\label{fig:1}
\end{figure*}

\begin{figure*}[ht!]
\centering
\includegraphics[width=0.32\textwidth]{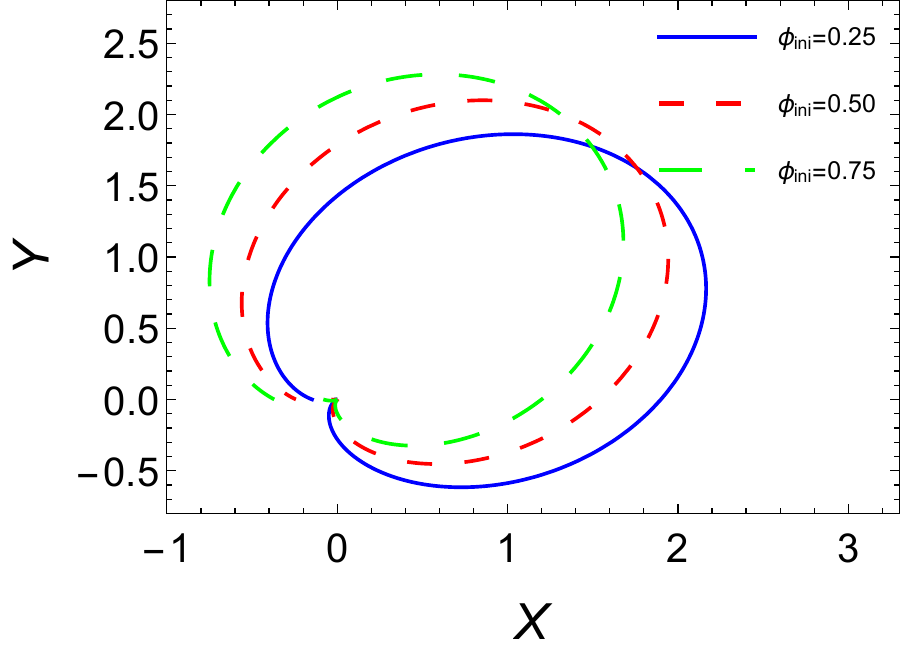}   \ \
\includegraphics[width=0.32\textwidth]{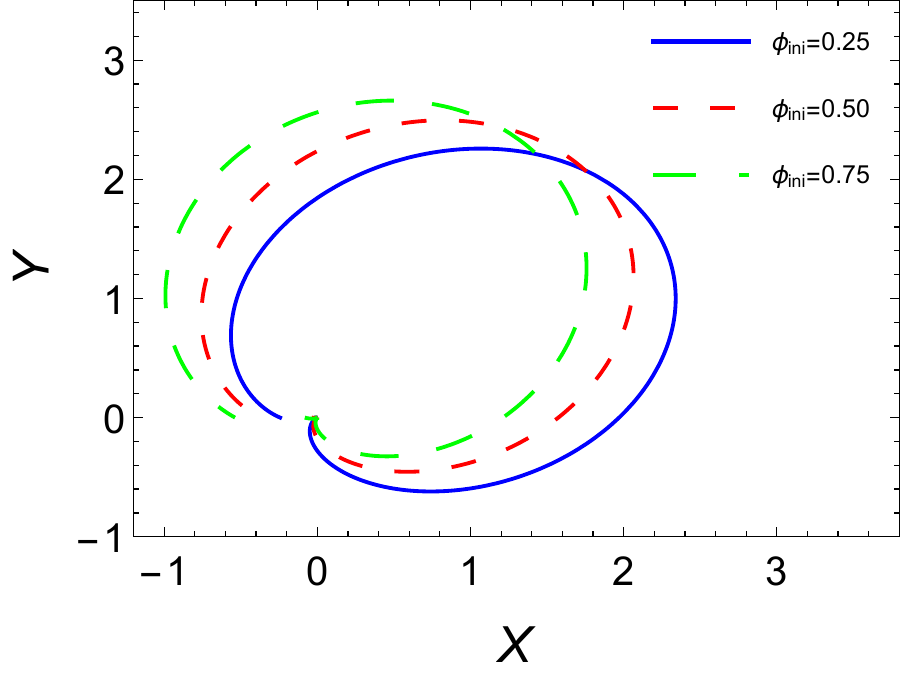}   \ \ 
\includegraphics[width=0.32\textwidth]{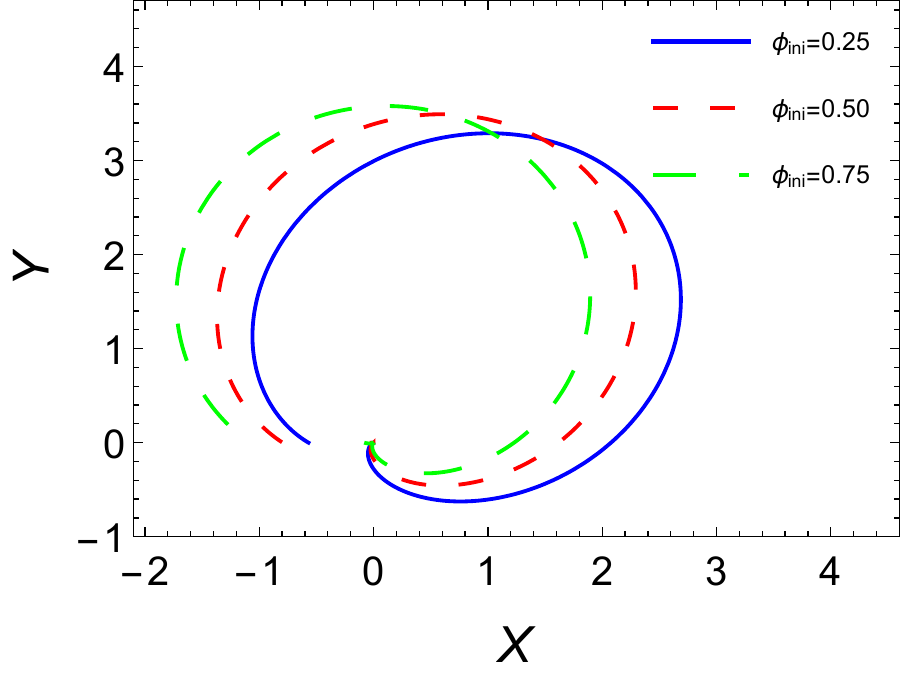}  % \ \
%\\
\includegraphics[width=0.32\textwidth]{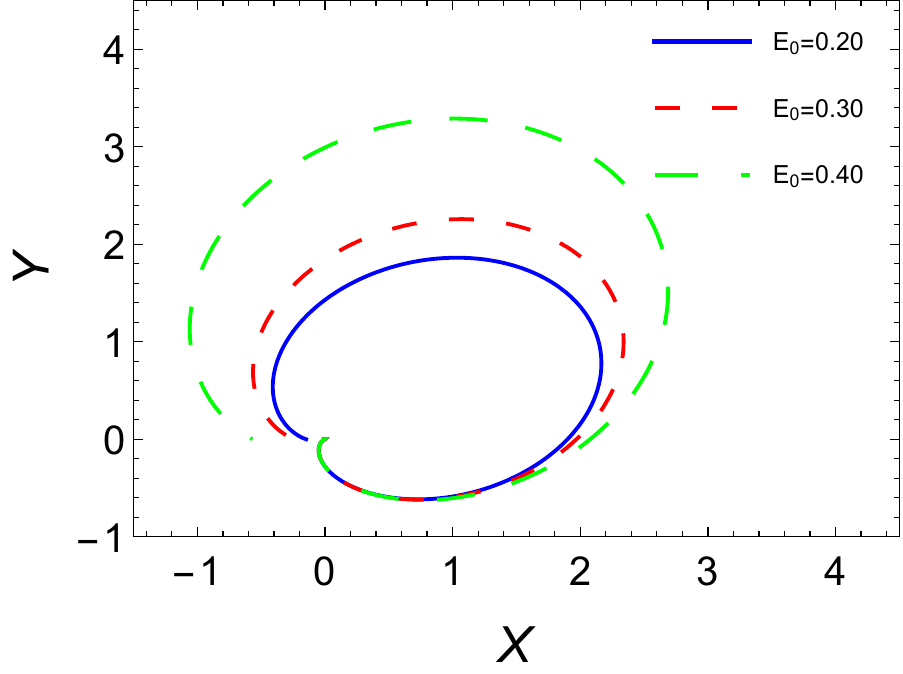}   \ \
\includegraphics[width=0.32\textwidth]{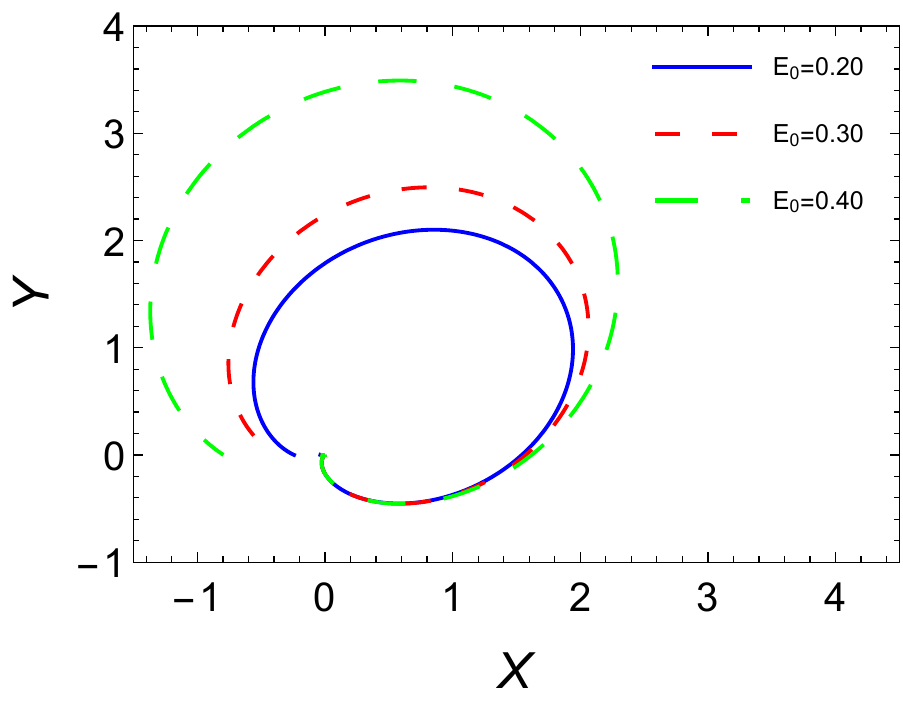}   \ \ 
\includegraphics[width=0.32\textwidth]{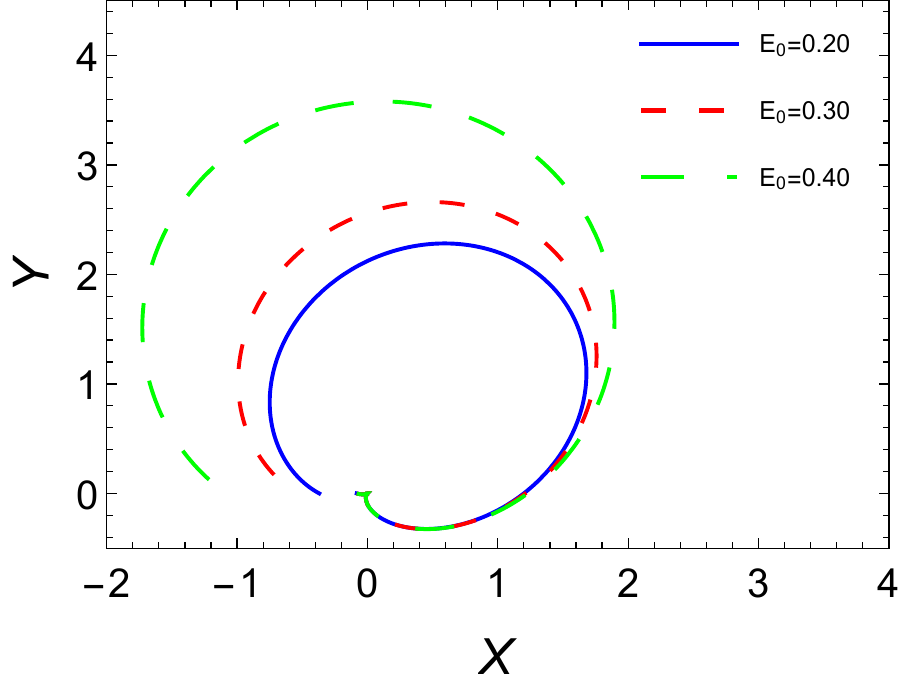}  % \ \
%\includegraphics[width=0.48\textwidth]{trayeSDE250v2.pdf}  
%}
\caption{ 
Photon orbits in the $(x-y)$ plane ($0 \leq \phi \leq 2 \pi$) in the Einstein-power-Maxwell geometry in (1+2) dimensions, setting $M=1, a=2, Q=0.5$ and $L=1$.
The impact of the photon energy and initial conditions is investigated
i) varying the initial angle for a given photon energy, and
ii) varying the energy for a given initial angle, 
as follows:
{\it{First row}}:
i) {\textbf{Left Panel:}} $E=0.20$ for different values of the initial angle $\phi_{ini}$.
ii) {\textbf{Middle Panel:}} $E=0.30$ for different values of the initial angle $\phi_{ini}$.
iii) {\textbf{Right Panel:}} $E=0.40$ for different values of the initial angle $\phi_{ini}$.
{\it{Second row}}:
i) {\textbf{Left Panel:}}  $\phi_{ini} = 0.25$  for different values of the energy $E$.
ii) {\textbf{Middle Panel:}} $\phi_{ini} = 0.50$   for different values of the energy.
iii) {\textbf{Right Panel:}} $\phi_{ini} = 0.75$  for different values of the energy.
}
\label{fig:2}
\end{figure*}

\begin{figure*}[ht!]
\centering
\includegraphics[width=0.32\textwidth]{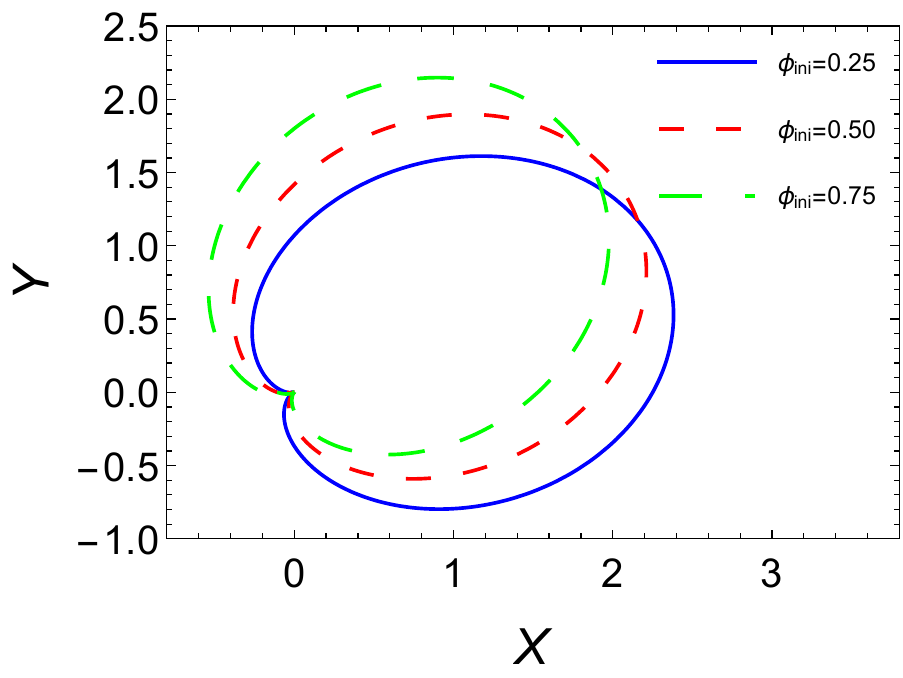}   \ \
\includegraphics[width=0.32\textwidth]{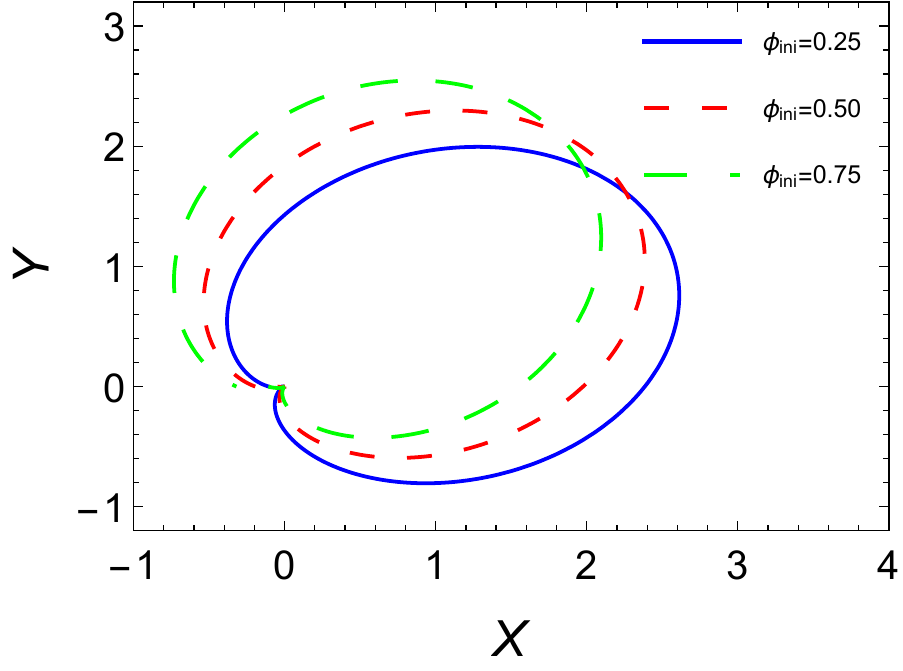}   \ \ 
\includegraphics[width=0.32\textwidth]{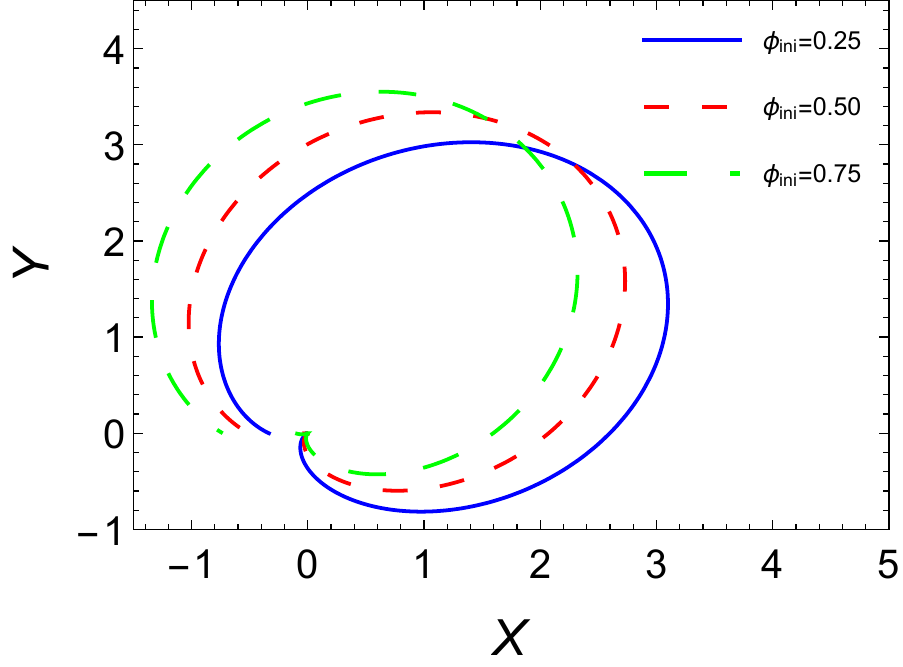}  % \ \
%\\
\includegraphics[width=0.32\textwidth]{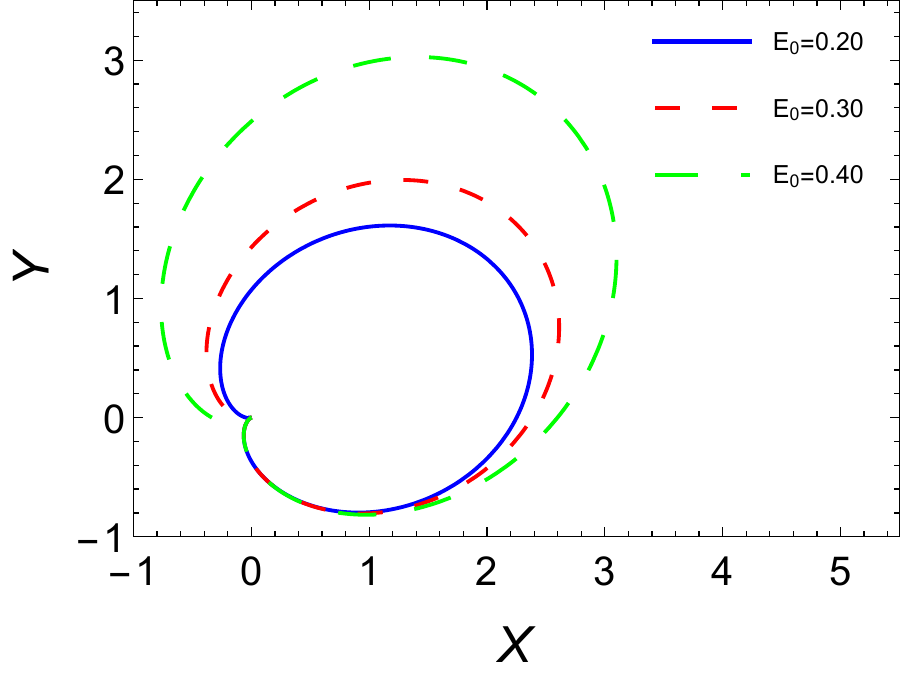}   \ \
\includegraphics[width=0.32\textwidth]{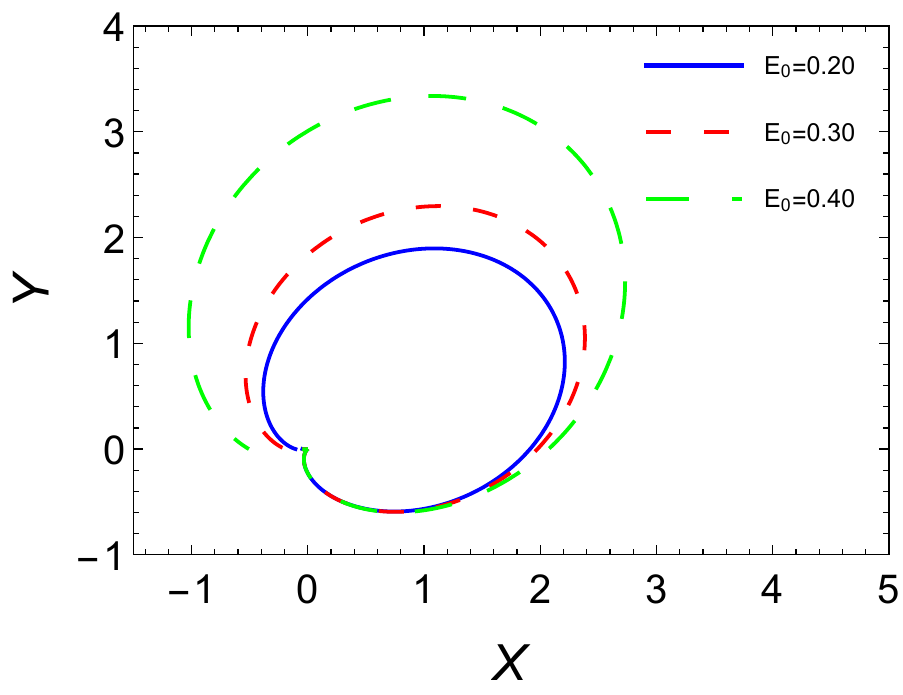}   \ \ 
\includegraphics[width=0.32\textwidth]{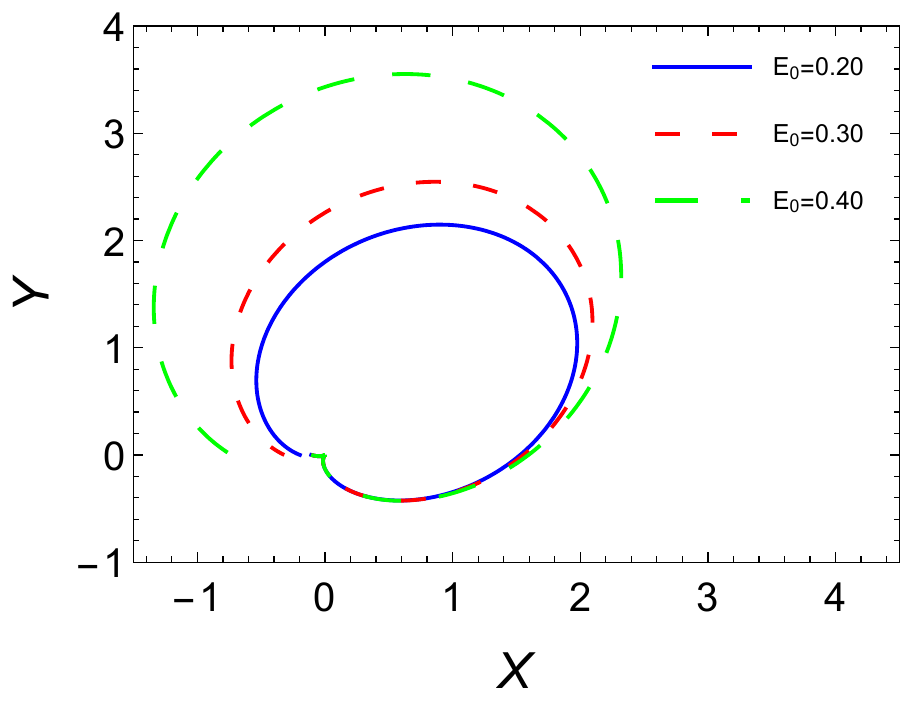}  % \ \
%\includegraphics[width=0.48\textwidth]{trayeSDE250v2.pdf}  
%}
\caption{
Photon orbits in the $(x-y)$ plane ($0 \leq \phi \leq 2 \pi$) in the Einstein-power-Maxwell geometry in (1+2) dimensions, setting $M=1, a=2, Q=0.6$ and $L=1$.
The impact of the photon energy and initial conditions is investigated
i) varying the initial angle for a given photon energy, and
ii) varying the energy for a given initial angle, 
as follows:
{\it{First row}}:
i) {\textbf{Left Panel:}} $E=0.20$ for different values of the initial angle $\phi_{ini}$.
ii) {\textbf{Middle Panel:}} $E=0.30$ for different values of the initial angle $\phi_{ini}$.
iii) {\textbf{Right Panel:}} $E=0.40$ for different values of the initial angle $\phi_{ini}$.
{\it{Second row}}:
i) {\textbf{Left Panel:}}  $\phi_{ini} = 0.25$  for different values of the energy $E$.
ii) {\textbf{Middle Panel:}} $\phi_{ini} = 0.50$   for different values of the energy.
iii) {\textbf{Right Panel:}} $\phi_{ini} = 0.75$  for different values of the energy.
}
\label{fig:3}
\end{figure*}

In Fig.~\eqref{fig:potentialscase1} we show the effective potential for the geometry investigated as well as three 
different values of the photon energy. Next, in Figures~\eqref{fig:1}, \eqref{fig:2} and \eqref{fig:3} we show the trajectories of the (1+2)-dimensional non-linear charged black hole, varying the photon energy, $E$, the initial 
angle, $\phi_{in},$ and the electric charge, $Q$.

\smallskip

We have considered three different values of the electric charge: i) $Q=0.4$, ii) $Q=0.5$ and iii) $Q=0.6$,
and three different values of the energy: 
i) $E=0.2$, ii) $E=0.3$ and iii) $E=0.4$. For each one of those, we have assumed three different values of the initial angle: i) $\phi_{in}=0.25$, ii) $\phi_{in}=0.5$ and iii) $\phi_{in}=0.75$.

\smallskip

We find terminating orbits at the singularity, where the initial point corresponds to the initial condition (depending on $\phi_{in}$) and the end point corresponds to the origin (0,0). Besides, variation of the initial angle rotates the orbit, while as the photon energy increases, the orbits occupy more space in the $X-Y$ plane. Finally, the orbits take the typical cardioid shape more and more as the electric charge increases.

%%%%%%%%%%%%%%%%%%FIGURES%%%%%%%%%%%%%%%%%%%%%%%%%%%%%%%%%%%%%%%%%%

%%%%%%%%%%%%%%%%%%%%%%%%%END FIGURES%%%%%%%%%%%%%%%%%%%%%%%%%%%%%%

%%%%%%%%%%%%%%%%%%%%%%%
\section{Conclusions}
%%%%%%%%%%%%%%%%%%%%%%

In the present work we have investigated the orbits of light rays in the gravitational background of $(1+2)$-dimensional 
Einstein-power-Maxwell black holes with circular symmetry within three-dimensional General Relativity.
We have obtained an exact analytic solution to the null geodesic equation in terms of the Weierstra{\ss} function.
The impact of the photon energy, the electric charge of the black hole, and the integration constant 
(initial condition) on the shape of the orbits has been investigated in detail. We find terminating orbits at the singularity. Moreover, our main numerical results may be summarized as follows: a) variation of the initial angle rotates the orbit, b) as the photon energy increases, the orbits occupy more space in the $X-Y$ plane, and c) the orbits take the typical cardioid shape more and more as the electric charge increases.

%%%%%%%%%%%%%%%%%%%%%%%%%%%
\section*{Acknowlegements}
%%%%%%%%%%%%%%%%%%%%%%%%%%%

A. R. is funded by the María Zambrano contract ZAMBRANO21-25 (Spain).

%%%%%%%%%%%%%%%%%%%%%%%%%%%%%%%%%%%%%%%%%%%%%%%%%%%%%%%%%%%%%%%%%%%%%%%%%%%%%%%%%

\end{document}